\documentclass[prd,preprint,nofootinbib]{revtex4}
\usepackage{amsmath}
\usepackage{graphicx}

\begin{document}
\begin{flushright}
CERN-PH-TH/2005-076
\end{flushright}
\title{Inclusive Production of a Higgs or $Z$ Boson \\ 
in Association with Heavy Quarks}
\author{Fabio Maltoni}
\affiliation{Theoretical Physics Division, CERN, 1211-CH Geneva,
Switzerland}
\author{Thomas McElmurry}
\author{Scott Willenbrock}
\affiliation{Department of Physics, University of Illinois at
Urbana-Champaign, 1110 West Green Street, Urbana, Illinois 61801}
\begin{abstract}
We calculate the cross section for the production of a $Z$ boson
in association with heavy quarks.  We suggest that this cross
section can be measured using an inclusive heavy-quark tagging
technique.  This could be used as a feasibility study for the
search for a Higgs boson produced in association with bottom
quarks.  We argue that the best formalism for calculating that
cross section is based on the leading-order process $b\bar b\to
h$, and that it is valid for all Higgs masses of interest at both
the Fermilab Tevatron and the CERN Large Hadron Collider.

\end{abstract}
\maketitle

\section{Introduction} \label{s:intro}
In the standard model, the Higgs boson has a very weak coupling to
bottom quarks. However, in a two-Higgs-doublet model, the coupling
of some or all of the physical Higgs particles to bottom quarks
can be greatly enhanced. For example, this occurs in the minimal
supersymmetric model for large values of $\tan\beta\equiv
v_2/v_1$, where $v_1$ and $v_2$ are the vacuum expectation values
of the Higgs doublets that couple to bottom and top quarks,
respectively. If the coupling is sufficiently enhanced, the
production of Higgs bosons in association with bottom quarks can
be an important process at the Fermilab Tevatron ($p\bar p$,
$\sqrt S = 1.96$ TeV) and the CERN Large Hadron Collider (LHC)
($pp$, $\sqrt S = 14$ TeV). A great deal of attention has been
directed towards this process
\cite{Dicus:1988cx,Kunszt:1991qe,Kao:1995gx,Dai:1994vu,Dai:1996rn,
Richter-Was:1997gi,Barger:1997pp,Choudhury:1998kr,Huang:1998vu,Drees:1997sh,
Carena:1998gk,Diaz-Cruz:1998qc,Balazs:1998nt,Balazs:1998sb,Dicus:1998hs,
unknown:1999fr,Carena:2000yx,Affolder:2000rg,Campbell:2002zm,Maltoni:2003pn,
Harlander:2003ai,Dittmaier:2003ej,Dawson:2003kb,Dawson:2004sh,Campbell:2004pu,
Abazov:2005yr}.

In order to separate the signal from the background, and also to
identify the production process, it is advantageous to tag one or
more of the bottom quarks produced along with the Higgs boson (in
addition to the bottom quarks that might result from Higgs decay).
Up until now, this has been discussed as the identification of a
high-$p_T$ $b$-tagged jet. However, there exist more inclusive
means to identify bottom quarks in the final state at hadron
colliders, such as identifying a secondary vertex without
requiring the reconstruction of a high-$p_T$ jet
\cite{Acosta:2004nj}. In this paper we would like to lay the
groundwork for such a measurement.

As a testing ground for the Higgs, we propose a measurement of the
inclusive production of a $Z$ boson in association with heavy
quarks.\footnote{The production of a $Z$ boson in association with
a heavy-quark jet is dealt with in
Refs.~\cite{Campbell:2003dd,Abazov:2004zd}.} This is more
complicated than the Higgs case for three reasons. First, the $Z$
boson is produced in association with both bottom and charm
quarks, so both possibilities must be taken into account. Second,
$Z$ bosons are dominantly produced in association with light
quarks, which can fake a heavy quark. Third, the processes $q\bar
q\to ZQ\overline Q$ and $qQ\to ZqQ$ ($Q=c,b$), where the $Z$
couples to the light quarks, are contributions that have no
analogue in the Higgs case.

There is a second motivation for this paper. There exist two
different formalisms for the calculation of inclusive Higgs
production in association with bottom quarks. The first is based
on the leading-order (LO) process $gg\to hb\bar b$, the second on
the LO process $b\bar b\to h$.\footnote{When a $b$ distribution
function is used, it is implicit that there is a spectator $\bar
b$ in the final state.} The advantage of the latter formalism is
that it resums, to all orders in perturbation theory, collinear
logarithms of the form $\ln(m_h/m_b)$ that arise in the
calculation based on $gg\to hb\bar b$
\cite{Aivazis:1993pi,Collins:1998rz}. It has recently been
suggested that both formalisms may be unreliable for Higgs bosons
at the Tevatron \cite{Kramer:2004ie}. We will show evidence that
the calculation based on $b\bar b\to h$ is reliable for Higgs
masses of experimental interest, and argue for its superiority.
However, we also find evidence that the formalism fails as the
Higgs mass approaches the machine energy, in agreement with
Ref.~\cite{Kramer:2004ie}.

The paper is organized as follows.   We first discuss, in
Section~\ref{s:h-prod}, the calculation of $b\bar b\to h$, and
argue that it is reliable for all Higgs masses of interest at the
Tevatron and LHC. We then turn in Section~\ref{s:Z-prod} to
inclusive production of a $Z$ boson with heavy quarks. Readers who
are only interested in the latter topic may skip directly to that
section. We conclude with a discussion of our results.

\section{Higgs production in association with heavy quarks} \label{s:h-prod}

Inclusive Higgs production in association with bottom quarks may
be calculated in two different schemes. One may work in a
four-flavor scheme, where the leading-order (LO) process is $gg\to
hb\bar b$. This approach yields collinear logarithms of the form
$\ln(m_h/m_b)$, which degrade the convergence of the perturbation
series. Alternatively, one may work in a five-flavor scheme, where
the LO process is $b\bar b\to h$
\cite{Aivazis:1993pi,Collins:1998rz}. The calculation based on
$b\bar b\to h$ yields a more convergent perturbation series, since
the collinear logarithms are summed into the $b$-quark
distribution functions via the
Dokshitzer-Gribov-Lipatov-Altarelli-Parisi (DGLAP) equations. As
one calculates to higher and higher order in perturbation theory,
the two calculations should approach each other, since they are
simply different orderings of the same terms.

The collinear logarithms that arise in $gg\to hb\bar b$ at LO can
be captured by an approximate $b$-quark distribution function,
\begin{equation*}
\tilde b(x,\mu_F)
=\frac{\alpha_S(\mu_F)}{2\pi}\ln\left(\frac{\mu_F^2}{m_b^2}\right)
\int_x^1\frac{dy}y\,P_{qg}\left(\frac xy\right)g(y,\mu_F)\;,
\end{equation*}
where $P_{qg}(x)=\frac12[x^2+(1-x)^2]$ is the LO DGLAP splitting
function and $\mu_F$ is the factorization scale, of order $m_h$.
Unlike the exact $b$ distribution function, the approximate $b$
distribution function does not sum the collinear logarithms. Thus
the calculations $gg\to hb\bar b$ and $\tilde b\bar{\tilde b}\to
h$ should approximately agree if the terms enhanced by collinear
logarithms in $gg\to hb\bar b$ are dominant.

Recently it was noted that the calculations $gg\to hb\bar b$ and
$\tilde b\bar{\tilde b}\to h$ differ substantially at LO for heavy
Higgs bosons ($m_h>100$ GeV) at the Tevatron, the discrepancy
increasing with increasing Higgs mass \cite{Kramer:2004ie}.  In
contrast, the two calculations agree fairly well at the LHC for
$m_h=100-500~\text{GeV}$ as well as at the Tevatron for $m_h<100$
GeV. Ref.~\cite{Kramer:2004ie} concludes that both calculations
are suspect at the Tevatron for $m_h>100$ GeV.

\begin{figure}
\begin{center}
\includegraphics[scale=0.35,angle=90]{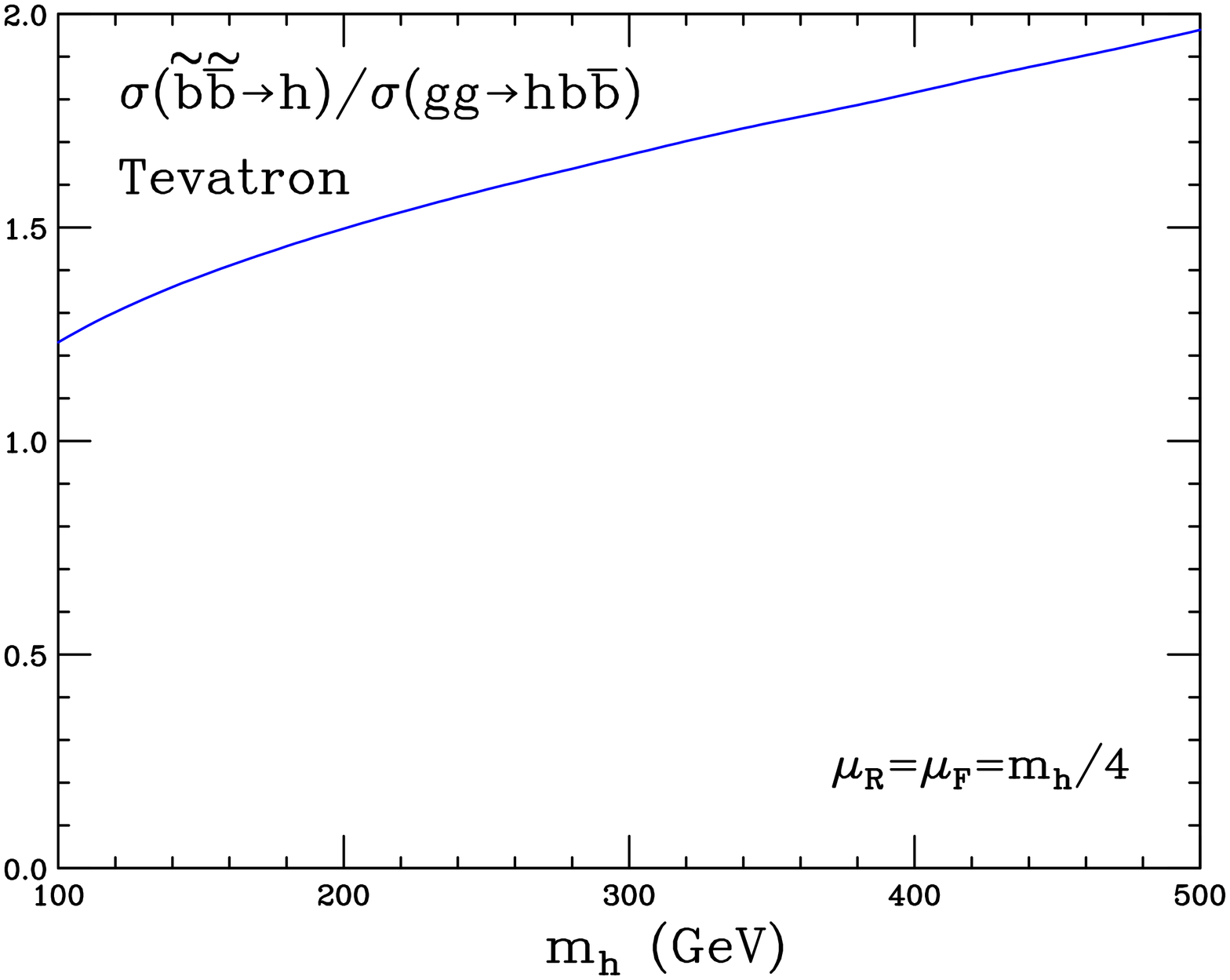}
\hspace{0.3in}
\includegraphics[scale=0.35,angle=90]{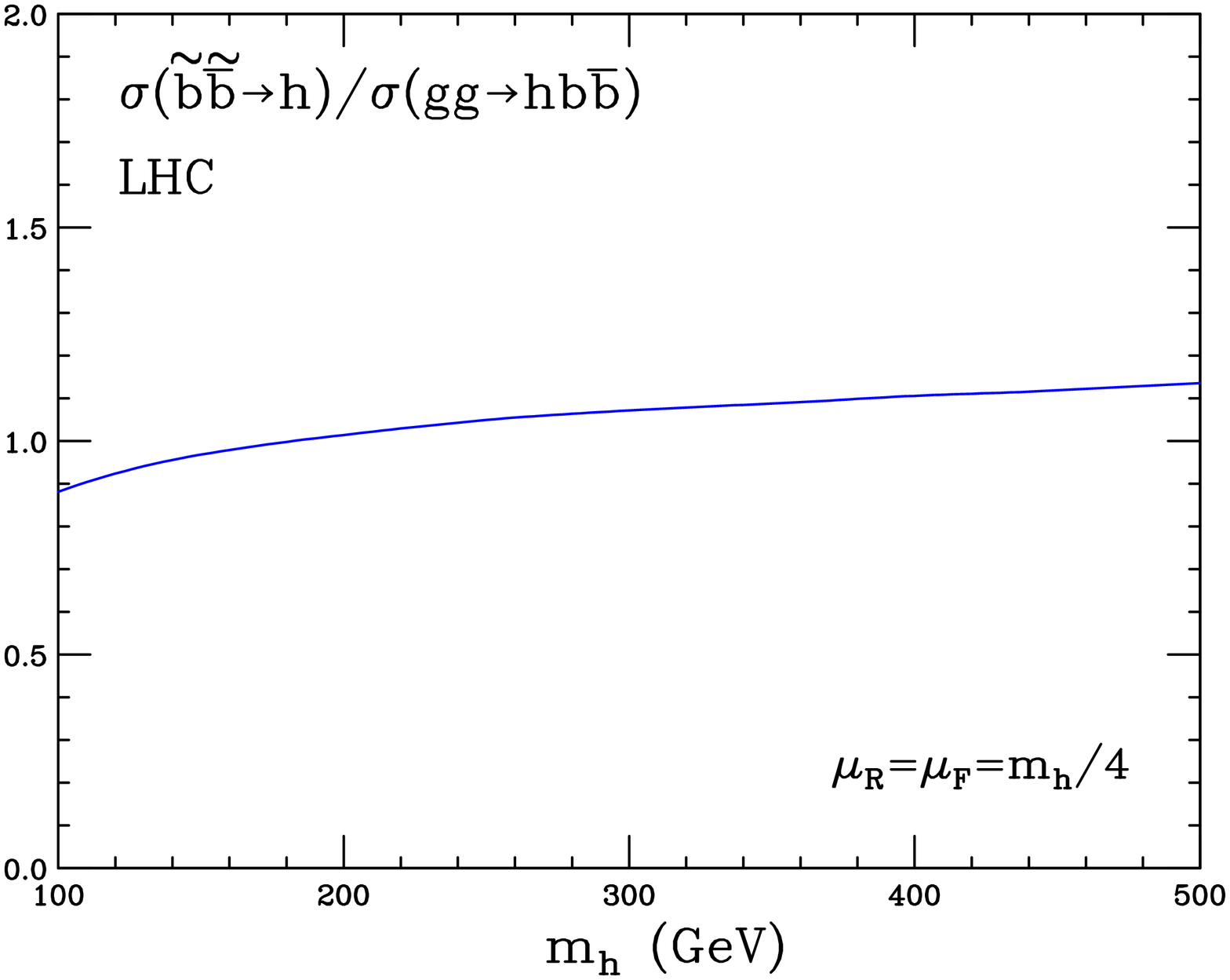}
\end{center}
\caption{$\sigma(\tilde b\bar{\tilde b}\to h)/\sigma(gg\to hb\bar
b)$ vs.\ $m_h$ at the Tevatron and the LHC, using MRST2001 LO
parton distribution functions \cite{Martin:2001es}, $m_b=4.7$ GeV,
and $\mu_F$ ($=\mu_R$) $=m_h/4$.} \label{f:ratio_fix}
\end{figure}

We show in Fig.~\ref{f:ratio_fix} the ratio of $\tilde
b\bar{\tilde b}\to h$ to $gg\to hb\bar b$ at both the Tevatron and
the LHC.  These results agree closely with those of
Ref.~\cite{Kramer:2004ie}.  We see that the ratio is about 1.5 for
$m_h=200$ GeV at the Tevatron, increasing to nearly 2 for
$m_h=500$ GeV.

Implicit in this argument is the choice of the factorization
scale. It was argued in Ref.~\cite{Maltoni:2003pn} that the
appropriate factorization scale is $\mu_F\approx m_h/4$, and this
is the scale that was used in Ref.~\cite{Kramer:2004ie} and
Fig.~\ref{f:ratio_fix}. We show below that for heavy Higgs bosons
at the Tevatron, a slightly lower scale is more appropriate, and
that this partially resolves the large discrepancy between the LO
calculations of $gg\to hb\bar b$ and $\tilde b\bar{\tilde b}\to
h$.

\begin{figure}
\begin{center}
\includegraphics[scale=0.35,angle=90]{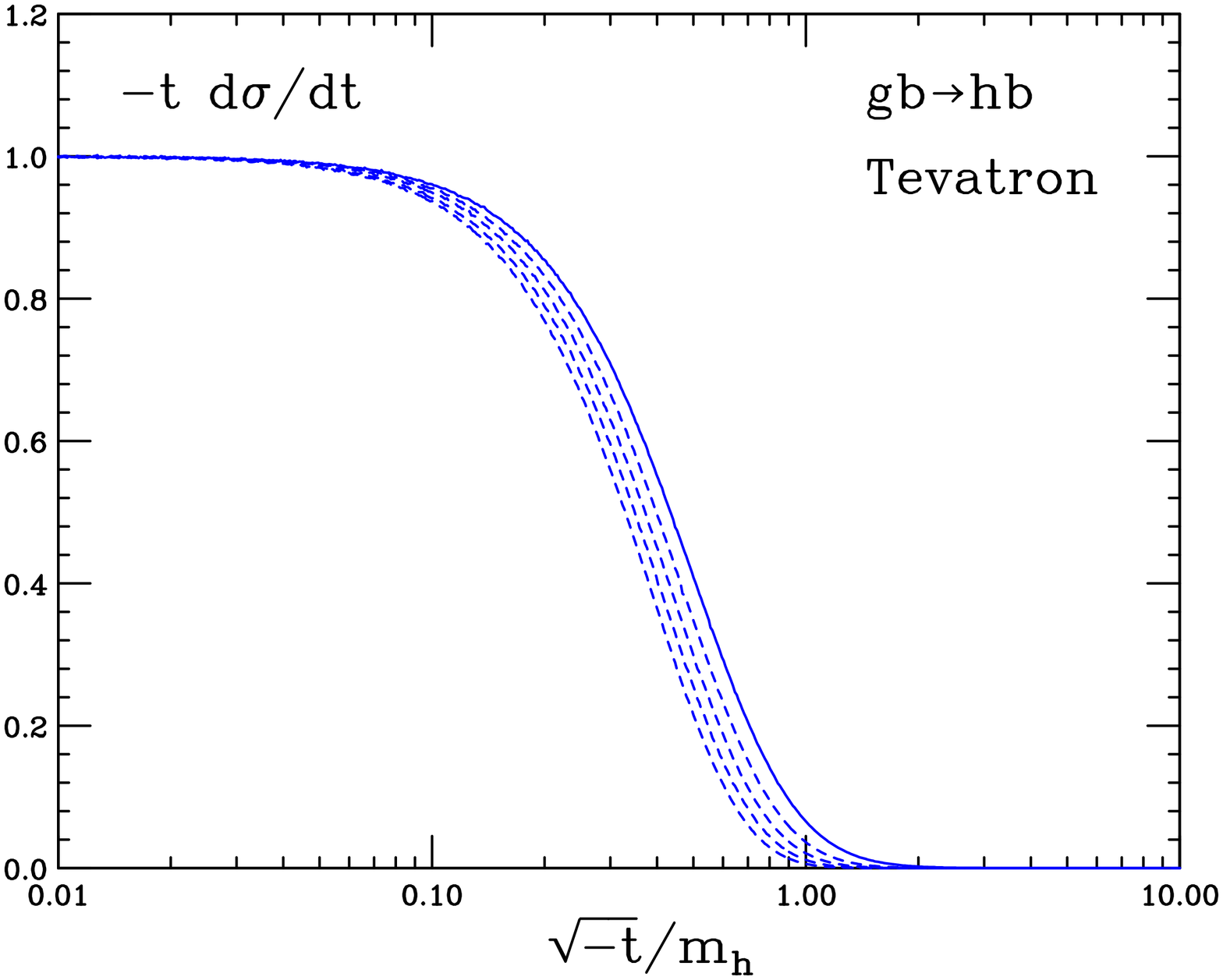}
\hspace{0.3in}
\includegraphics[scale=0.35,angle=90]{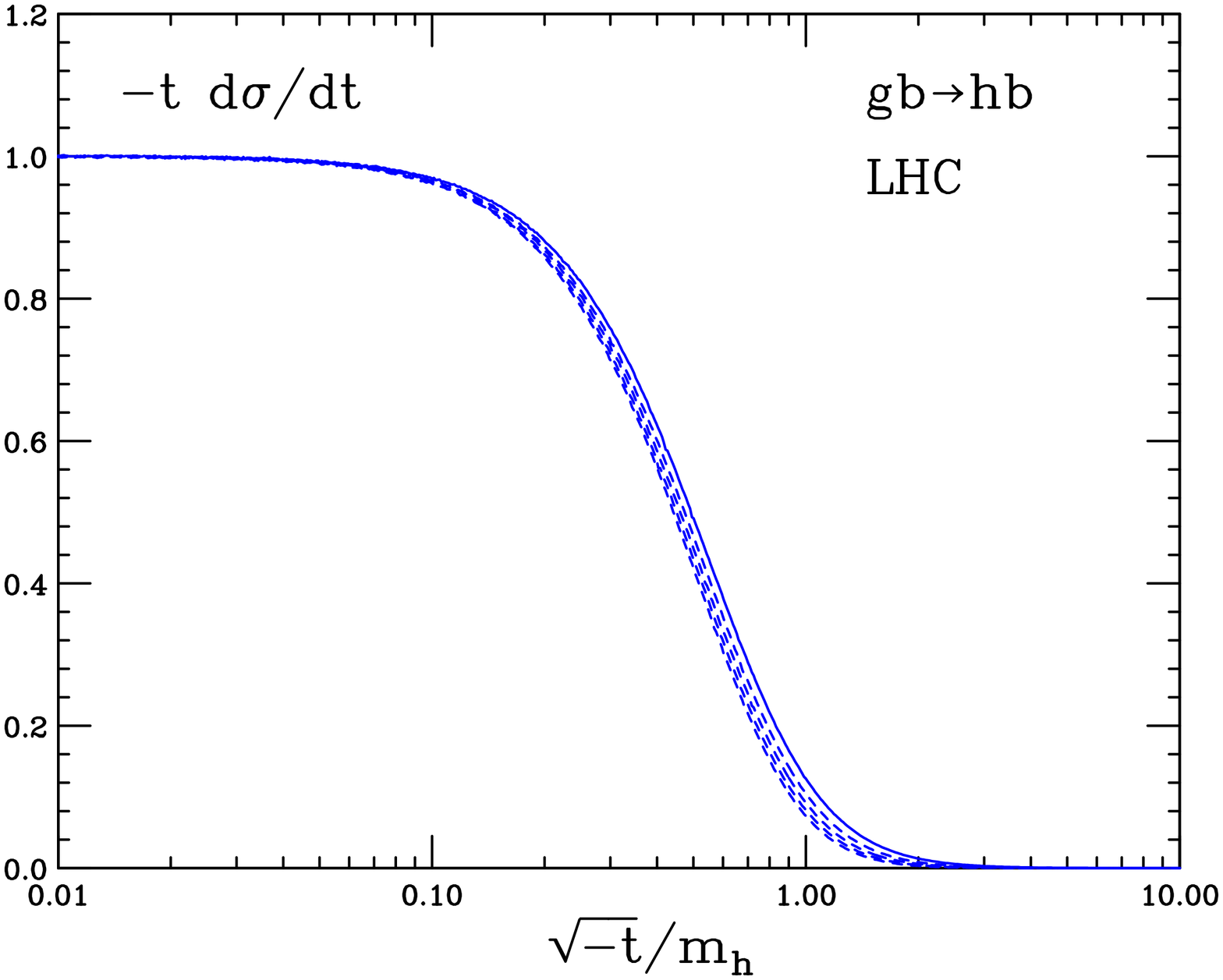}
\end{center}
\caption{$-t\,d\sigma/dt$ vs.\ $\sqrt{-t}/m_h$ for $gb\to hb$ at
the Tevatron and the LHC.  The factorization scale for $b\bar b\to
h$ should be chosen near the end of the collinear plateau.}
\label{f:plateau}
\end{figure}

The argument for the factorization scale made in
Ref.~\cite{Maltoni:2003pn} is based on an analysis of the
collinear logarithm that arises at next-to-leading order (NLO) in
the calculation of $b\bar b\to h$, and is in the same spirit as
the argument of Refs.~\cite{Plehn:2002vy,Boos:2003yi}. In the
collinear region, the NLO differential hadronic cross section
scales like $d\sigma/dt\sim 1/t$, where $t$ is the usual
Mandelstam variable. We show in Fig.~\ref{f:plateau} the quantity
$-t\,d\sigma/dt$ vs.\ $\sqrt{-t}/m_h$ for the NLO process $gb\to
hb$ at both the Tevatron and the LHC for $m_h=100-500$
GeV.\footnote{Since this is a NLO process, we use NLO parton
distribution functions \cite{Martin:2002dr}. We use $\mu_F=m_h$
(the default value), as this graph is being used to determine
$\mu_F$. We subsequently check that the curves are not very
sensitive to the choice of $\mu_F$.} The factorization scale
should be chosen near the end of the collinear plateau. At the LHC
this plateau ends around $m_h/4$ for the Higgs-boson masses
considered. However, at the Tevatron the end of the plateau slowly
creeps below $m_h/4$ as the Higgs-boson mass increases (this is
also true at the LHC, but much less so).

\begin{figure}
\begin{center}
\includegraphics[scale=0.35,angle=90]{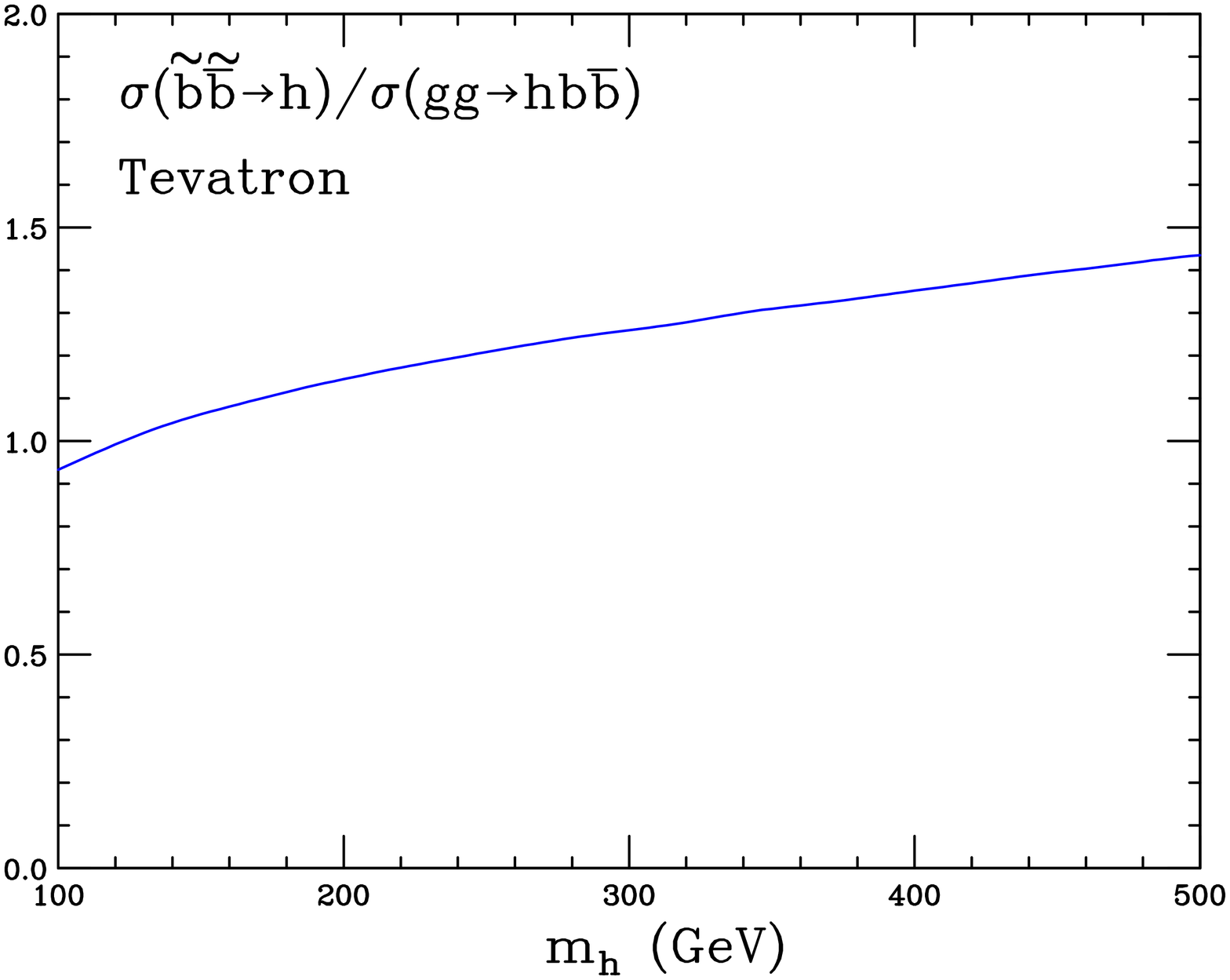}
\hspace{0.3in}
\includegraphics[scale=0.35,angle=90]{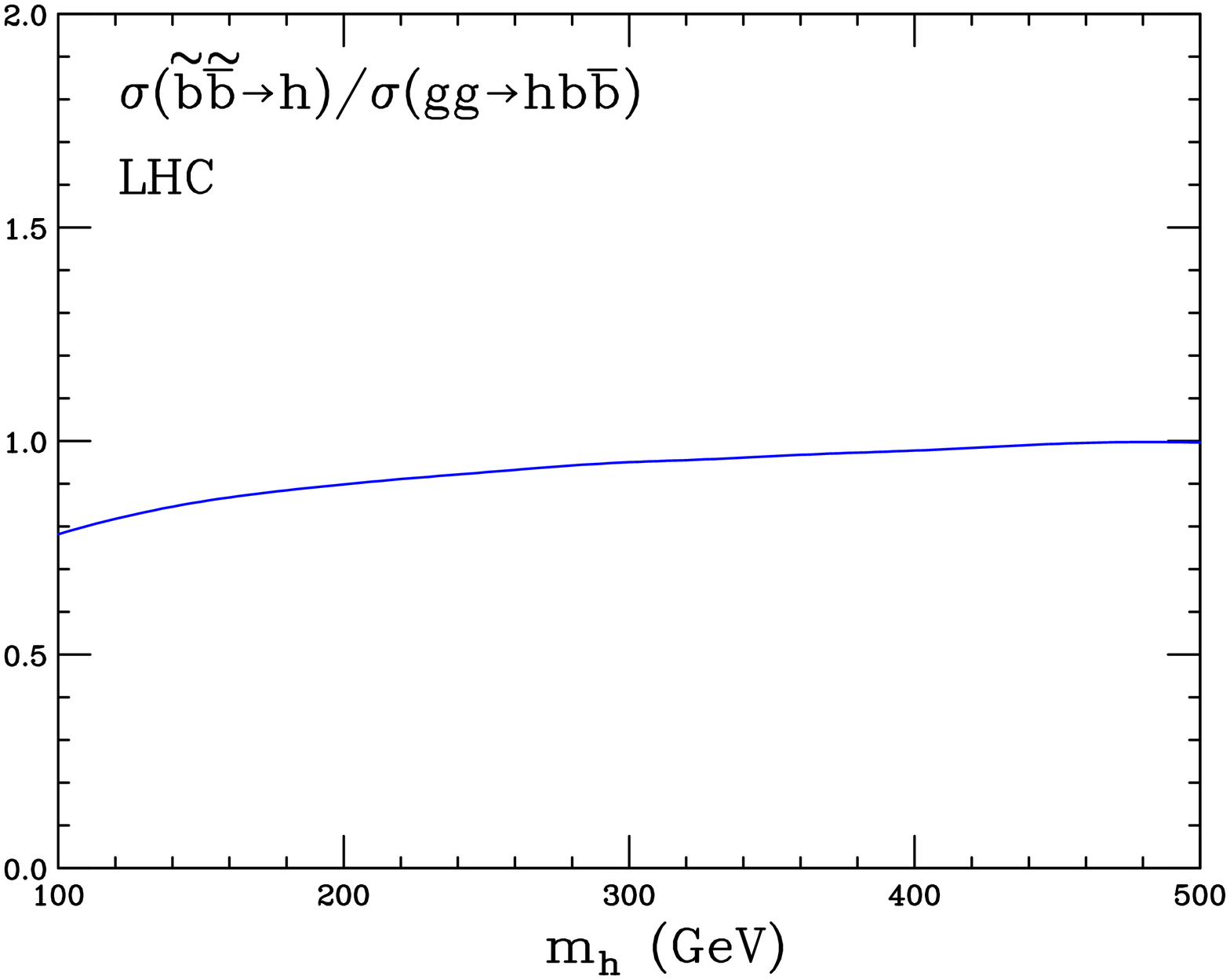}
\end{center}
\caption{$\sigma(\tilde b\bar{\tilde b}\to h)/\sigma(gg\to hb\bar
b)$ vs.\ $m_h$ at the Tevatron and the LHC, using MRST2001 LO
parton distribution functions \cite{Martin:2001es}, $m_b=4.7$ GeV,
and $\mu_F$ ($=\mu_R$) determined from the end of the collinear
plateau in Fig.~\ref{f:plateau} (listed in Table~\ref{t:scale}).}
\label{f:ratio}
\end{figure}

To be consistent, we choose the factorization scale to be where
$-t\,d\sigma/dt$ reaches 85\% of its value on the collinear
plateau. The resulting factorization scale at both the Tevatron
and the LHC is given in Table~\ref{t:scale}.  We show in
Fig.~\ref{f:ratio} the ratio of $\tilde b\bar{\tilde b}\to h$ to
$gg\to hb\bar b$ at both the Tevatron and the LHC with this choice
of factorization scale. The ratio approaches unity for large Higgs
masses at the LHC, as would be expected if the collinear
logarithms dominate.  The situation at the Tevatron is more
complicated.  The ratio is near unity for Higgs masses of
experimental interest, indicating that the calculation is
reliable.  However, as the Higgs mass increases the ratio grows,
and continues to grow as the mass approaches the machine energy.
This suggests that the calculation based on $b\bar b\to h$ may be
unreliable for very heavy Higgs bosons at the Tevatron.

\begin{table}
\begin{center}\begin{tabular}{|c|c|c|}
\hline
$m_h$ [GeV] & Tevatron & LHC \\
\hline
100 & 0.203 & 0.227 \\
200 & 0.188 & 0.219 \\
300 & 0.176 & 0.215 \\
400 & 0.166 & 0.210 \\
500 & 0.157 & 0.206 \\
\hline
\end{tabular}\end{center}
\caption{The factorization scale relative to the Higgs mass,
$\mu_F/m_h$, at the Tevatron and the LHC.  The factorization scale
is determined by the point at which the curves in
Fig.~\ref{f:plateau} reach 85\% of their values on the collinear
plateau.} \label{t:scale}
\end{table}

In the full calculation of $b\bar b\to h$ (using the exact $b$
distribution function) it is important to choose the factorization
scale near the end of the collinear plateau, but not very
important exactly how that is defined.  A less-than-optimal choice
will be corrected by higher orders.  Indeed, the
next-to-next-to-leading-order (NNLO) calculation of $b\bar b\to h$
has very little factorization-scale dependence for values of
$\mu_F$ near the end of the collinear plateau
\cite{Harlander:2003ai}.

The advantage of the calculation based on the LO process $b\bar
b\to h$ is actually twofold. As already discussed, it gives a more
convergent perturbation series. In addition, it allows for a
higher-order calculation than $gg\to hb\bar b$, since it is a
simpler LO process. Indeed, $b\bar b\to h$ is known at NNLO
\cite{Harlander:2003ai}, while $gg\to hb\bar b$ is known only at
NLO \cite{Dittmaier:2003ej,Dawson:2003kb,Dawson:2004sh}. Thus the
NNLO calculation of $b\bar b\to h$ is the most accurate existing
calculation of inclusive Higgs-boson production in association
with bottom quarks. This is reflected by the very mild dependence
of the NNLO calculation of $b\bar b\to h$ on the factorization
scale in comparison with that of the NLO calculation of $gg\to
hb\bar b$ \cite{Campbell:2004pu}.  It would be interesting to
study the behavior of the NNLO calculation for very heavy Higgs
bosons at the Tevatron.

\section{$Z$ production in association with heavy quarks} \label{s:Z-prod}

\begin{figure}
\begin{center}
\includegraphics[scale=0.8,angle=0]{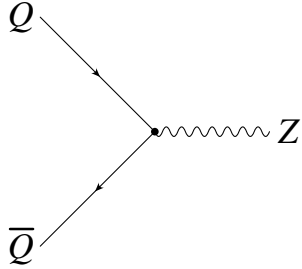}
\end{center}
\caption{Feynman diagram for $Q\bar Q\to Z$ ($Q=c,b$). The
presence of heavy quarks in the final state is implied by the
initial-state heavy quarks.} \label{f:QQZ}
\end{figure}

\begin{figure}
\begin{center}
\includegraphics[scale=0.8,angle=0]{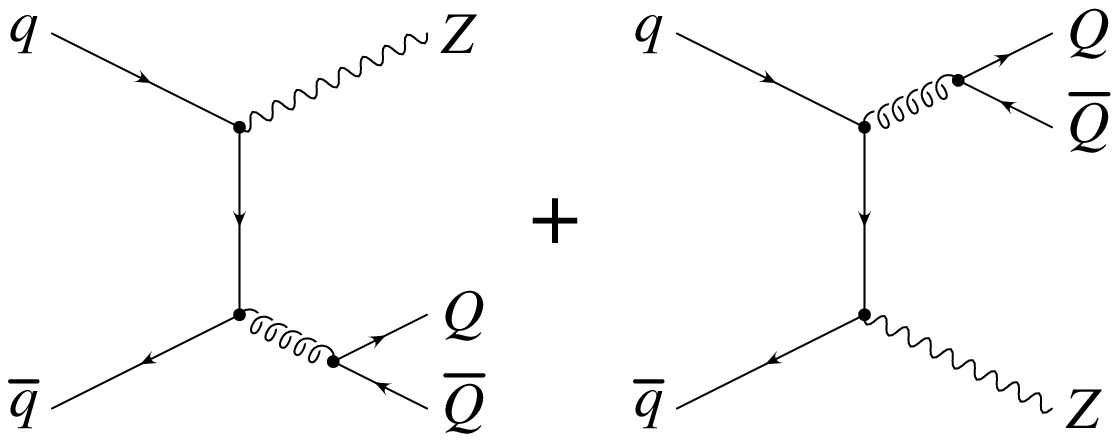}
\end{center}
\caption{Feynman diagrams for $q\bar q\to ZQ\overline Q$, where
the $Z$ couples to the light quarks.} \label{f:qqZQQ}
\end{figure}

\begin{figure}
\begin{center}
\includegraphics[scale=0.8,angle=0]{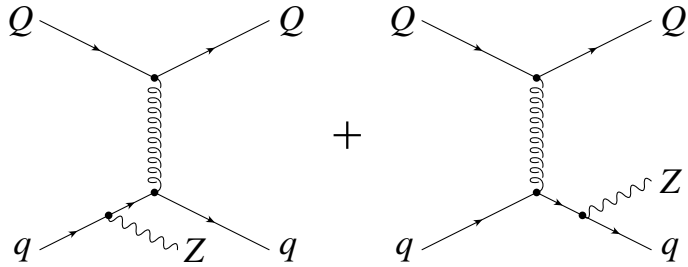}
\end{center}
\caption{Feynman diagrams for $qQ\to ZqQ$, where the $Z$ couples
to the light quarks} \label{f:qQqQZ}
\end{figure}

Unlike the case of the Higgs boson, there are a variety of
contributions to the inclusive production of a $Z$ boson with
heavy quarks.  The analogue of the Higgs case is $b\bar b\to Z$,
shown in Fig.~\ref{f:QQZ}. In the case of the $Z$ boson, one must
also consider $c\bar c\to Z$ and $q\bar q\to Z$ ($q=u,d,s$), since
both charm quarks and light quarks can fake a $b$ quark. Finally,
there are the processes $q\bar q\to ZQ\overline Q$ and $qQ\to ZqQ$
($Q=c,b$), shown in Figs.~\ref{f:qqZQQ} and \ref{f:qQqQZ}, where
the $Z$ boson couples to the light quarks. As we will show, these
last two processes are more important at the Tevatron than at the
LHC.

Let us begin by considering only processes in which the $Z$ boson
couples to the heavy quarks.  This is completely analogous to the
case of the Higgs boson discussed in the previous section.  We
will then include processes in which the $Z$ boson couples to
light quarks, which have no analogue in the Higgs case.

\begin{figure}
\begin{center}
\includegraphics[scale=0.35,angle=90]{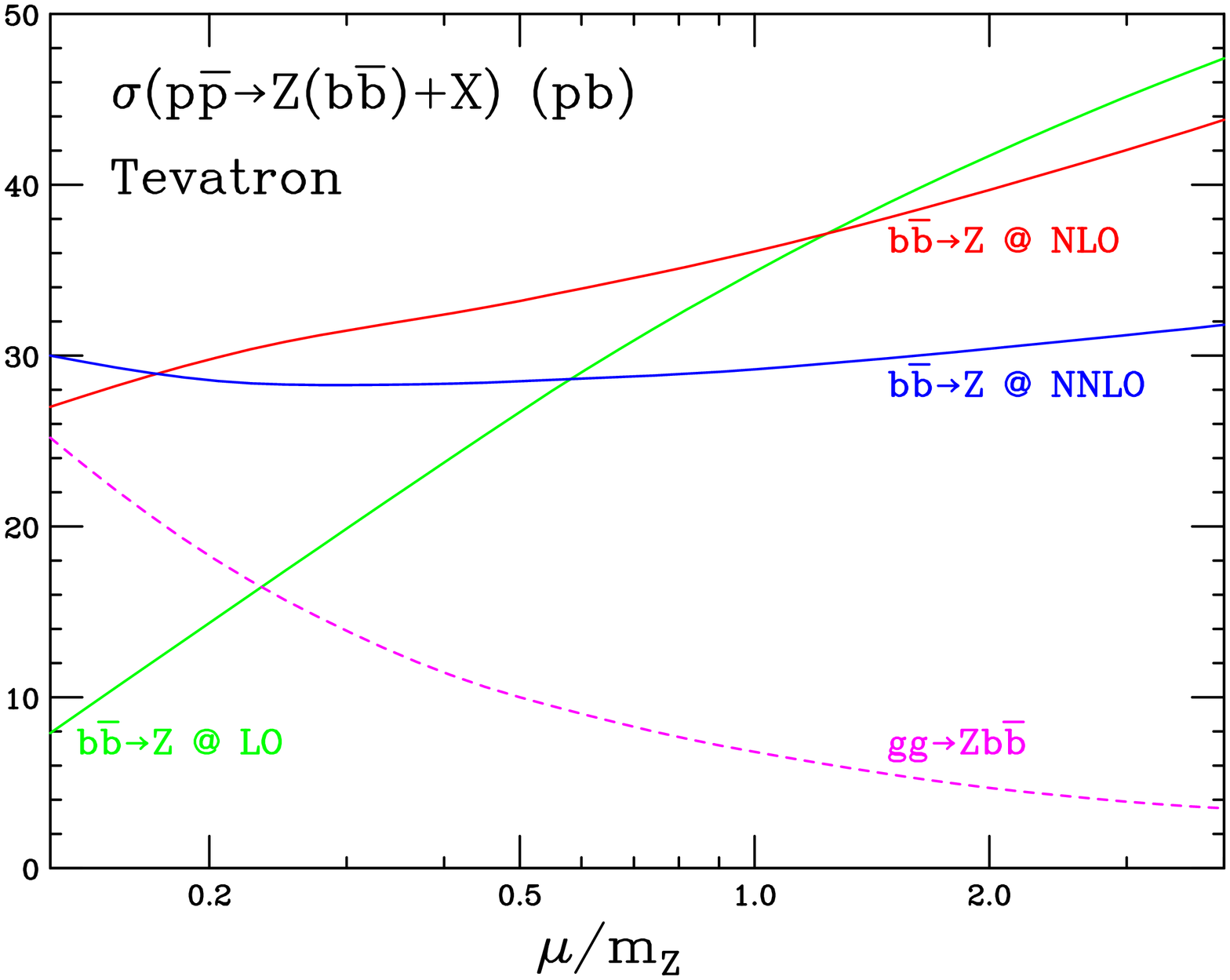}
\hspace{0.2in}
\includegraphics[scale=0.35,angle=90]{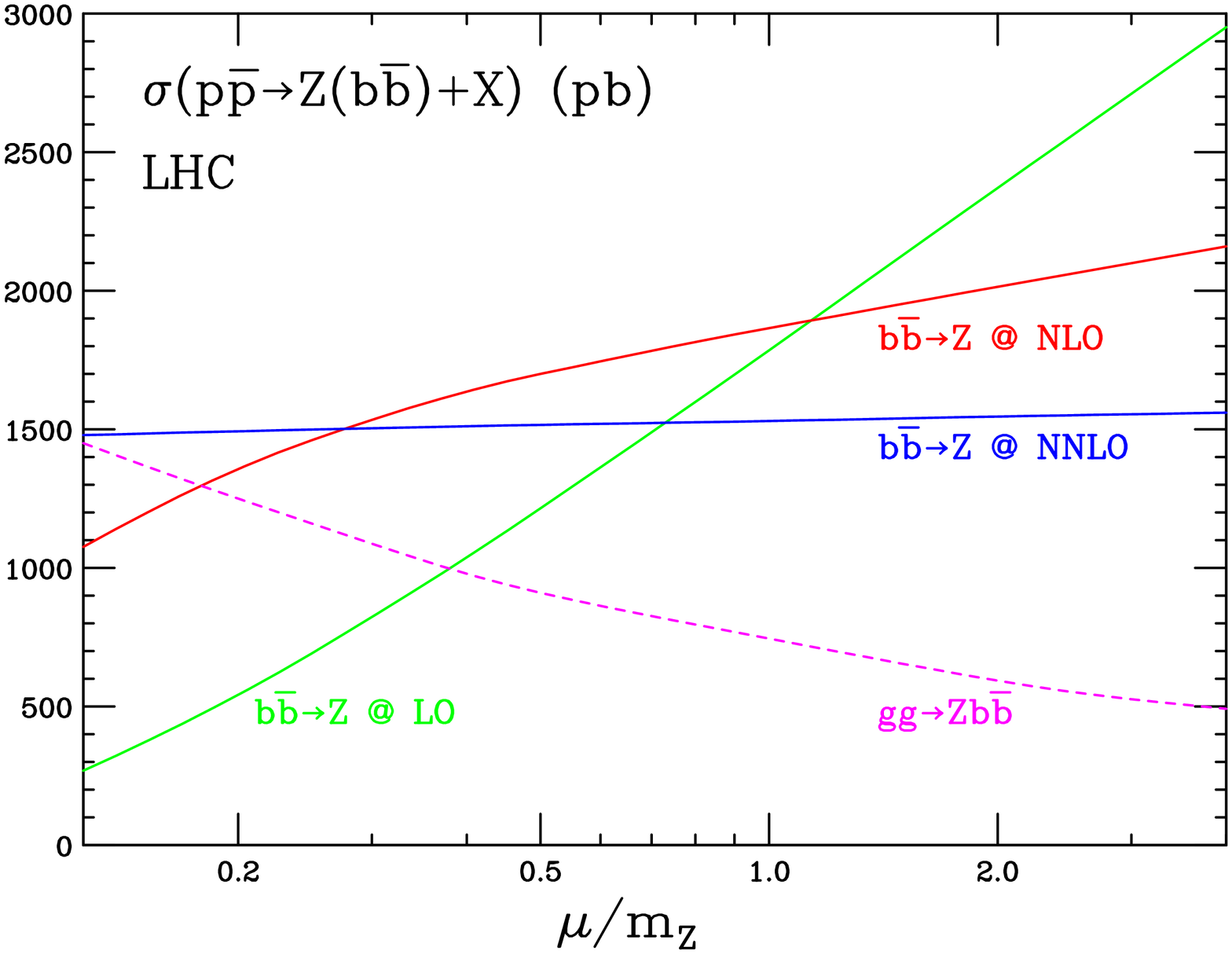}
\end{center}
\caption{Factorization-scale dependence of $b\bar b\to Z$ at LO,
NLO, and NNLO at the Tevatron and the LHC. Only processes in which
the $Z$ couples to the heavy quarks are included.  Also shown is
$gg\to Zb\bar b$ at LO, using $m_b=4.7$ GeV.  We use the LO, NLO,
and NNLO parton distribution functions MRST2001 LO
\cite{Martin:2001es} and MRST2002 \cite{Martin:2002dr}.}
\label{f:Zbbmu}
\end{figure}

\begin{figure}
\begin{center}
\includegraphics[scale=0.35,angle=90]{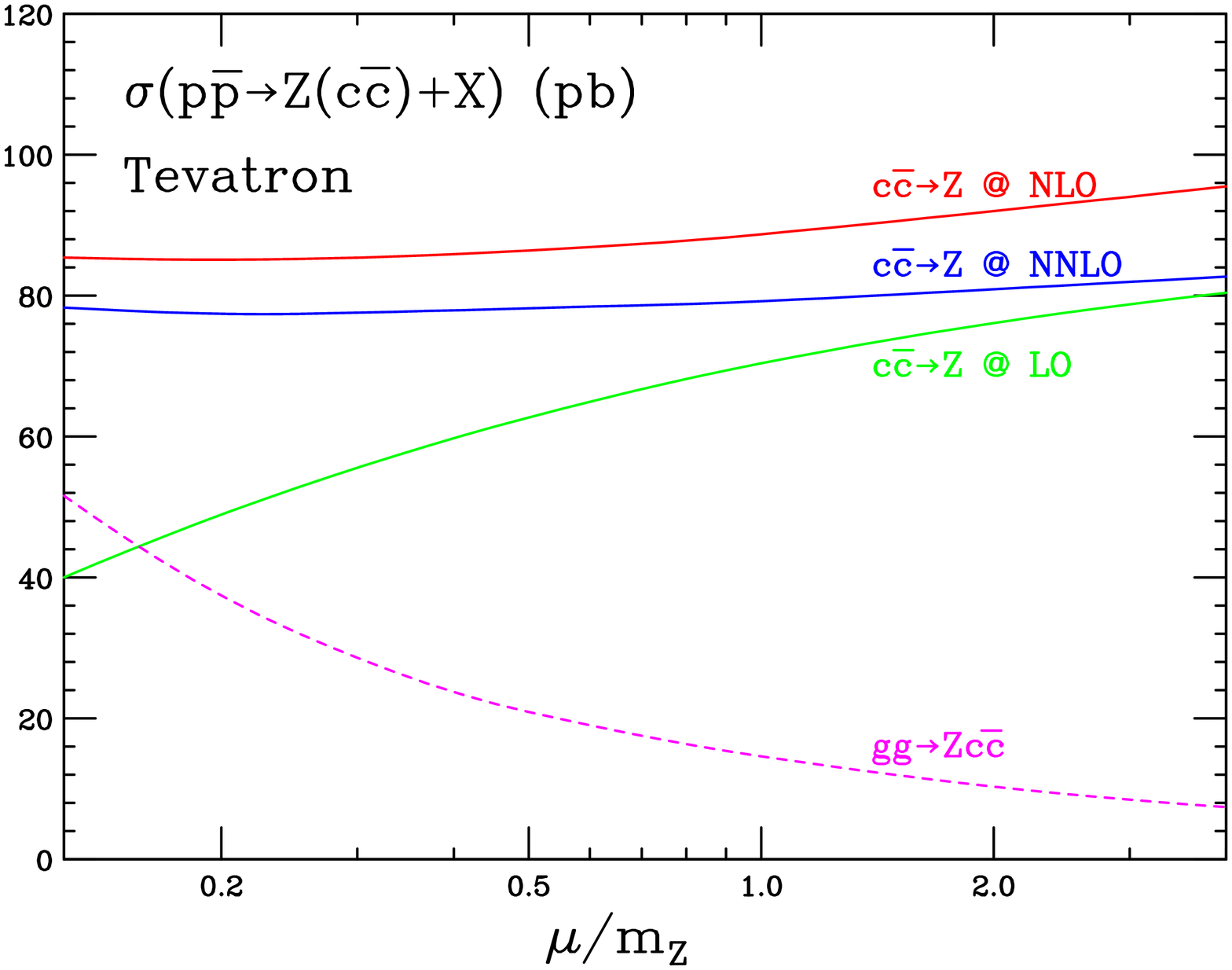}
\hspace{0.2in}
\includegraphics[scale=0.35,angle=90]{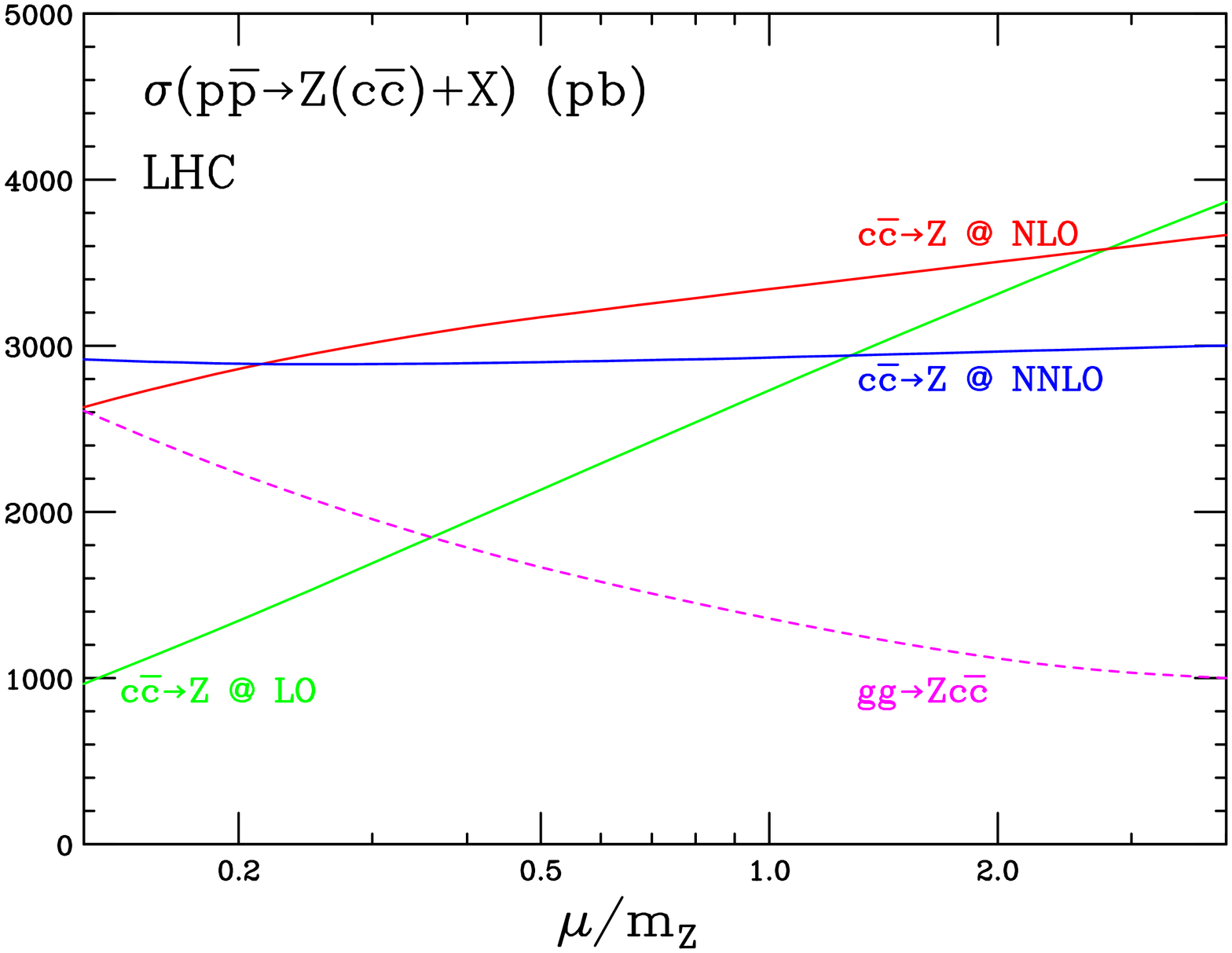}
\end{center}
\caption{Same as Fig.~\ref{f:Zbbmu}, but for $c\bar c\to Z$.  For
$gg\to Zc\bar c$, we use $m_c=1.4$ GeV.} \label{f:Zccmu}
\end{figure}

The process $q\bar q\to Z$ has been calculated at
next-to-next-to-leading order (NNLO)
\cite{Hamberg:1990np,Rijken:1995gi,Harlander:2002wh}. We modified
this code to extract $Q\overline Q\to Z$ ($Q=c,b$) at NNLO,
neglecting the heavy-quark mass, which is a small effect of order
$(m_Q/M_Z)^2\times 1/\ln^2(M_Z/m_Q)$.  We keep (for the moment)
only diagrams in which the $Z$ couples to the heavy
quarks.\footnote{This includes NNLO processes with four external
heavy quarks of the same flavor. However, we do not include
processes with two external charm and two external bottom quarks.
These processes contribute less than 1\% of the LO cross section.}
We show in Fig.~\ref{f:Zbbmu} the factorization-scale dependence
of the cross section for $b\bar b\to Z$ at both the Tevatron and
the LHC at LO, NLO, and NNLO. The renormalization scale has been
set equal to the factorization scale, although this hardly matters
as it first enters only at NLO, via the argument of
$\alpha_S(\mu_R)$. As expected, the scale dependence decreases
with increasing order, to the point where there is almost no scale
dependence at NNLO. Similar results are obtained for $c\bar c\to
Z$, as shown in Fig.~\ref{f:Zccmu}.

Also shown on the same plot is the LO cross section in the
four-flavor scheme, $gg\to Zb\bar b$.  If we were to choose
$\mu_F=\mu_R=M_Z$, this calculation would underestimate the true
cross section by a factor of 4 at the Tevatron.  If we choose the
scale as in the previous section, by finding the end of the
collinear plateau in $gb\to Zb$, we find that the appropriate
factorization scale is around $M_Z/3$ at both the Tevatron and the
LHC.  With this choice of scale, the factor of 4 is reduced to 2.
For charm, the corresponding factor is 5 at the Tevatron, reduced
to 3. These results mirror a similar result that was obtained in
the case of the Higgs \cite{Maltoni:2003pn}.

We list in Table~\ref{t:cross-sections} the NNLO cross sections
for $b\bar b\to Z$ and  $c\bar c\to Z$ at both the Tevatron and
the LHC.  These cross sections have very little theoretical
uncertainty.

We now include processes in which the $Z$ couples to light quarks,
shown in Figs.~\ref{f:qqZQQ} and \ref{f:qQqQZ}.\footnote{We also
consider the process $Q\bar Q\to Zq\bar q$, where the $Z$ couples
to the light quarks. We find this to be numerically negligible.
The interference with the same process, but where the $Z$ couples
to the heavy quarks, is also negligible.} Here it is essential to
keep the heavy-quark mass nonzero in order to regulate collinear
singularities.\footnote{We also evaluate the interference of these
processes with the similar processes in which the $Z$ is radiated
from the heavy quark.  This may be done in the limit of vanishing
heavy-quark mass, since there are no collinear singularities.  We
find that these interference contributions are numerically
negligible.} While these processes are NNLO with respect to
inclusive $Z$ production, $q\bar q\to ZQ\overline Q$ is LO with
respect to $Z$ production in association with heavy quarks, and
$qQ\to ZqQ$ is NLO.  The correct power counting is obtained when
one recalls that a heavy-quark distribution function is
intrinsically of order $\alpha_S\ln(\mu_F/m_Q)$
\cite{Stelzer:1997ns}. The analogous processes for heavy-quark
structure functions in deep-inelastic scattering, $F_i^Q$, have
been discussed in Ref.~\cite{Chuvakin:1999nx}.

There are two serious drawbacks to the calculations of these
processes.  First, the cross sections contain factors of $\ln
(M_Z/m_Q)$, due to the collinear singularities, which are not
resummed.  This is related to the fact that we are calculating a
semi-inclusive quantity, namely $Z$ production in association with
heavy quarks.  If we were instead calculating the inclusive $Z$
cross section, this issue would not arise.  Fracture functions may
be useful in this context \cite{Trentadue:1993ka}. Second, and
more importantly, these processes are only known at LO at this
time (with a nonzero quark mass).\footnote{The process $q\bar q\to
ZQ\bar Q$ is known at NLO with a vanishing quark mass, which is
relevant when the heavy quarks are produced at high $p_T$
\cite{Campbell:2000bg}.} The NLO calculation is an important
missing result for this as well as many other analyses (the same
holds true of $q\bar q\to WQ\overline Q$). Thus our calculation of
these processes is relatively crude. This is a serious problem at
the Tevatron, but less so at the LHC, where these processes are
relatively less important.  It is desirable both to obtain NLO
results for $q\bar q\to ZQ\bar Q$ and $qQ\to ZqQ$ (with finite
$m_Q$) and to develop a formalism that allows the resummation of
the collinear logarithms.\footnote{The collinear logarithm that
occurs in $qQ\to ZqQ$ also occurs in the process $qg\to ZqQ\bar
Q$, where the initial gluon splits to $Q\bar Q$.  Thus the same
issues arise in a calculation that does not make use of a
heavy-quark distribution function.}

We give in Table~\ref{t:cross-sections} the cross sections for the
various processes that contribute to $Z$ production in association
with heavy quarks.  We also give the inclusive $Z$ cross section
at NNLO. Although this cross section is two orders of magnitude
larger than that of any of the processes that produce a $Z$ in
association with heavy quarks, the mistag rate for light quarks
and gluons is on the order of 1\%, so this background is not
overwhelming.

\begin{table}
\label{tab:2} \addtolength{\arraycolsep}{0.1cm}
\renewcommand{\arraystretch}{1.4}
\begin{center}
\begin{tabular}{|c|l|c|c|}
\hline
\multicolumn{2}{|c|}{Process} & Tevatron & LHC \\
\hline
                 &$b \bar b \to Z$ (NNLO)         & 28.3 & 1500  \\
$Z (b \bar b)$ &$q\bar q  \to Z b \bar b$ (LO)  &   19 &  120  \\
                 &$q b      \to Z q b$ (LO)       &  5.9 &  430  \\
\hline
                 &$c \bar c \to Z$ (NNLO)         & 77.7 & 2890\\
$Z (c \bar c)$ &$q\bar q  \to Z c \bar c$ (LO)  & 69   & 430 \\
                 &$q c   \to Z q c $ (LO)         & 21   & 1200\\
\hline
\multicolumn{2}{|c|}{Inclusive $Z$}                 & 7510 & 56700\\
\hline
\end{tabular}
\end{center}
\caption{Cross sections (pb) for the various contributions to $Z$
production in association with heavy quarks.  We use the MRST2002
NNLO parton distribution functions \cite{Martin:2002dr} with
$\mu_F=\mu_R=M_Z/3$.} \label{t:cross-sections}
\end{table}

\section{Conclusions}

Our results for $Z$ production in association with heavy quarks
are summarized in Table~\ref{t:cross-sections}.  The final row is
the inclusive $Z$ cross section calculated at
next-to-next-to-leading order (NNLO).  Each row above that
corresponds to some subset of this calculation, with heavy quarks
in the final state, either implicitly (such as $Q\bar Q\to Z$) or
explicitly (such as $q\bar q\to ZQ\bar Q$). We see that there are
a large variety of processes that contribute to $Z$ production in
association with heavy quarks. Taken together, they constitute 3\%
of the inclusive $Z$ cross section at the Tevatron, and 12\% at
the LHC.  The measurement of these fractions will require
simulation of the acceptances and tagging efficiencies of the
various processes.  We advocate using an inclusive tagging
technique to maximize the number of signal events.

The measurement of $Z$ production in association with heavy quarks
is interesting in its own right, but also as a feasibility study
for Higgs production in association with bottom quarks.  In this
paper we have argued that the best formalism for the calculation
of the latter process is based on the leading-order process $b\bar
b\to h$, and that this calculation is valid for all Higgs masses
of interest at both the Tevatron and the LHC.  The most accurate
calculation of this process is the NNLO cross section given in
Ref.~\cite{Harlander:2003ai}.  We also showed evidence that this
formalism fails as the Higgs mass approaches the machine energy.

\section*{Acknowledgements}

We are grateful for conversations about theoretical issues with
J.~Collins, S.~Forte, and F.~Olness, and about experimental issues
with T.~Liss, S.~Marcellini, G.~Masetti, S.~Nikitenko, and
K.~Pitts. This work was supported in part by the U.~S.~Department
of Energy under contract No.~DE-FG02-91ER40677.

\end{document}